# Resonant TMR inversion in LiF/EuS based spin-filter tunnel junctions

Fen Liu[1,2], Yihang Yang[2], Qian Xue[2], Zhiwei Gao[2], Aixi Chen[1,2], Guo-Xing Miao[2*]

1. East China Jiaotong University, Department of Science, Nanchang, Jiangxi 330013, PR China

2. Institute for Quantum Computing & Department of Electrical and Computer Engineering, University of Waterloo, Waterloo, ON N2L 3G1 Canada

**ABSTRACT:**

Resonant tunneling can lead to inverse tunnel magnetoresistance when impurity levels rather than direct tunneling dominate the transport process. We fabricated hybrid magnetic tunnel junctions of CoFe/LiF/EuS/Ti, with an epitaxial LiF energy barrier joined with a polycrystalline EuS spin-filter barrier. Due to the water solubility of LiF, the devices were fully packaged *in situ*. The devices showed sizeable positive TMR up to 16% at low bias voltages but clearly inverted TMR at higher bias voltages. The TMR inversion depends sensitively on the thickness of LiF, and the tendency of inversion disappears when LiF gets thick enough and recovers its intrinsic properties.

**Introduction:**

A traditional Magnetic tunnel junction consists of two ferromagnetic layers (electrodes) separated by a thin insulator barrier, such as MgO or $Al_2O_3$. When the relative alignment between the electrodes' magnetization changes, the net conductivity across the junction changes as well, generating the desired tunneling magnetoresistance (TMR) effect [1]. The TMR technology has already been widely used in spintronic devices such as magnetic sensors and magnetic random access memories [2]. The TMR ratio is commonly defined as,

$$\text{TMR} = \frac{R_{AP} - R_P}{R_{AP}}$$

where $R_P$ and $R_{AP}$ are the resistance of the tunnel junction with parallel (P) and antiparallel (AP) magnetic alignments, respectively. In a standard TMR device, we would expect the AP configuration to have higher resistance because of the spin misalignment, therefore the TMR ratio is positive in most cases. However, as many theoretical and experimental studies have already explored, the sign of TMR can be affected by many factors: 1) the intrinsic spin polarization of certain ferromagnets being minority spin dominated, such as $Fe_3O_4$ [3]; ii) the mobility of *sp-* or *d*-electrons being different, rendering the dominant spins in transport different from the dominant spins populating the Fermi level, such as Co coupled with $Al_2O_3$ barriers [4]; iii) the effect of bonding near the Fermi level at the metal-oxide interface, such as that between Co and $SrTiO_3$ [5]; iv) resonant tunneling due to the presence of impurity levels inside the tunnel barrier [6]. Among these mechanisms, the impurity mediated resonant inversion is clearly extrinsic and varies with sample preparation. In this article, we studied a hybrid tunnel barrier, LiF/EuS, combining spins from a conventional ferromagnetic electrode (CoFe) with a filtered spin current from EuS to produce TMR. The native TMR is positive in this system, as revealed at low bias voltages. On the other hand, LiF is a very reactive material prone to impurities and defects [7], and we clearly observed inverted TMR at higher than certain threshold voltages. As the LiF thickness increases, the threshold voltages increase and eventually disappear, indicating a gradual perfection of the barrier material itself.

LiF has a rock-salt structure very similar to that of MgO, therefore it can be effectively coupled to bcc (body-centered-cubic) ferromagnets and potentially induce large symmetry-assisted spin polarization [8]. Due to the high ion mobility in LiF, defects readily form during the deposition process, which become the main source of resonance tunneling centers. On the other hand, EuS is a well-studied spin-filter material with quite significant spin-filtering efficiency [9,10]. Because the tunnel barrier exchange splits into a lower one and a higher one for the two spin channels, different spins will tunnel with drastically different probabilities. An unpolarized current therefore turns into a highly polarized spin current after crossing the spin-filter barrier. Joining a ferromagnetic electrode with a spin-filter, one can generate TMR without needing the second ferromagnetic electrode. Similar spin-filter tunnel junctions have been reported with $Al_2O_3$ [11,12] and MgO [13] as the nonmagnetic barriers. Note that no TMR reversal has been observed in those devices because $Al_2O_3$ and MgO are some of the "perfect" barrier materials. We therefore turn our attention to LiF, and try to identify its role in generating TMR inversion.

**Experiment:**

In this work, the tunnel junctions have the basic structure (in nm): 4 $Co_{50}Fe_{50}$ / $x$ LiF / 3 EuS / 10 Ti. Because of the water solubility of LiF, the whole device structure was created with an *in situ* shadow mask technique instead of lithography. The samples were prepared on HF etched Si(100) substrates. We deposited all the films via electron beam evaporation in a high vacuum chamber with the base pressure about $2\times10^{-9}$ torr. The typical deposition rates were all maintained at approximately 0.1 Å/s with the pressure less than $1\times10^{-8}$ torr during the depositions. In order to seed epitaxial growth, a 10 nm MgO film was first laid down as the buffer layer with the substrate heated at 300 °C. The sample was then cooled down to near room temperature before depositing CoFe and LiF/EuS, in order to minimize the potential pinhole formation in the barrier. Due to the water solubility of LiF, the active junction areas (30×30 µm$^2$) were *in situ* patterned with shadow evaporations to avoid wet lithography procedures. This was achieved with two cross "definition" steps: each step places a set of laser-machined stainless steel bars (30 µm width) covering the bottom electrodes, and evaporates 5 nm MgO everywhere else leaving only the 30 µm opening as the active area. We then laid down 10 nm Ti as the top electrode. The samples were stored in dry nitrogen until measurement. The transport measurement was done in liquid helium bath at 4 K.

**Results and Discussion:**

Through x-ray diffraction (XRD), we can verify that both CoFe and LiF are epitaxial on the MgO buffer layer. Fig.1 shows the ω-2θ scan on a thicker sample stack. EuS is clearly polycrystalline from the appearance of both (200) and (111) peaks. On the other hand, MgO, CoFe, and LiF are all in the (200) orientation. We then performed an off-axis Φ scan with the sample tilted by 54.7°, targeting the LiF{111} reflections. The clear 4-fold symmetry verifies that LiF is indeed epitaxial, which also proves that the underlying MgO and CoFe are both epitaxial. From similar Φ scans targeting the MgO{220} and CoFe{110} reflections (where Si{220} peaks also showed up due to the extremely high intensity), we can deduce that MgO cells are sitting straight on top of Si cells, CoFe in-plane rotated 45° relative to MgO, and LiF again in-plane rotated 45° relative to MgO (therefore straight with respect to the substrate). With only 0.2% lattice mismatch between LiF and CoFe, the epitaxy of LiF on CoFe is straightforward. Epitaxy of EuS on LiF would be possible with their 1.2% mismatch (assuming 3:2 stacking), but the extremely high mobility/reactivity of LiF prevents us from applying higher tempera-tures to promote epitaxy. Instead, we had to limit LiF from being exposed to higher temperatures in order to create more stable tunnel junctions. As such, EuS was deposited at near room temperature and turned out not epitaxial in our junctions.

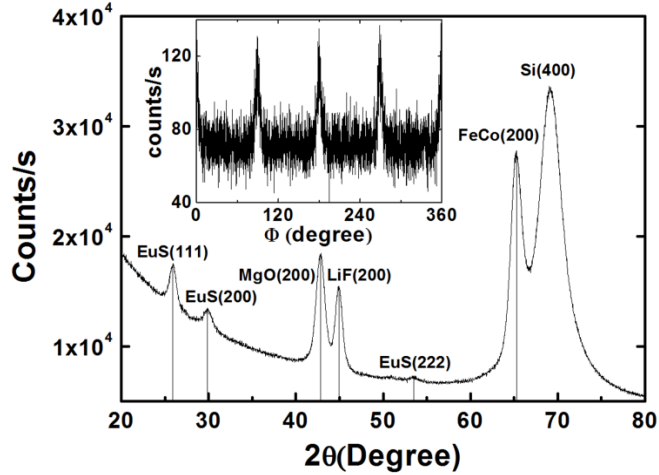

Fig.1: ω-2θ X-ray diffraction on a stack of (in nm): 10 MgO / 10 $Co_{50}Fe_{50}$ / 10 LiF / 10 EuS. Here the angle ω was offset from θ by 2° to avoid the extremely strong Si(400) and its satellite peaks. The inset shows the off-axis Φ scan targeting the LiF {111} peaks, with the 4-fold symmetry proving that LiF is epitaxial. Clearly the sample stays epitaxial till LiF, and then EuS becomes polycrystalline.

The intrinsic TMR of these devices are all positive, which can be seen from the TMR loops at small bias voltages, such as that in Fig.2. This is because the intrinsic spin polarization from the $Co_{50}Fe_{50}$ $\Delta_1$ bands, and the intrinsic spin-filtering efficiency from EuS, are both positive, therefore a positive TMR ratio is naturally expected. As the bias voltage increases, however, some other conduction channels are activated. As a result, the TMR magnitude, sometimes even its polarity, also changes. Fig.2 inset shows an example of the same junction measured at a much higher bias voltage. The TMR ratio is clearly inverted, with the antiparallel state showing lower resistance. This turned out to be a generic feature of all our junctions. Since EuS is clearly not the source of TMR inversion as evidenced from the literatures, we focus our attention to the new material introduced - LiF. In Fig.3 we summarize the TMR bias dependence for different LiF thicknesses, while keeping the EuS thickness at a constant 3 nm. We note that when we increase the LiF layer thickness, the voltage required to invert TMR becomes larger and larger, and the amplitude of inversion gets smaller and smaller. On the negative voltage branch (defined as electrons flow from the Ti electrode towards CoFe), the inversion point is 0.17 V for 0.8 nm LiF, 0.27 V for 0.9 nm, 0.48 V for 1 nm, and no inversion for 1.1 nm. For positive bias voltages, the inversion voltage is 0.46 V for 0.8 nm LiF, 0.5 V for 0.9 nm, 0.59 V for 1 nm, and 0.99 V for 1.1 nm. Note that even though inversion still happens for 1.1 nm LiF on the positive branch, the tendency of inversion is clearly diminished. The asymmetric behavior is attributed to the asymmetric junction itself, whether the electrons travelling from EuS into LiF or vice versa.

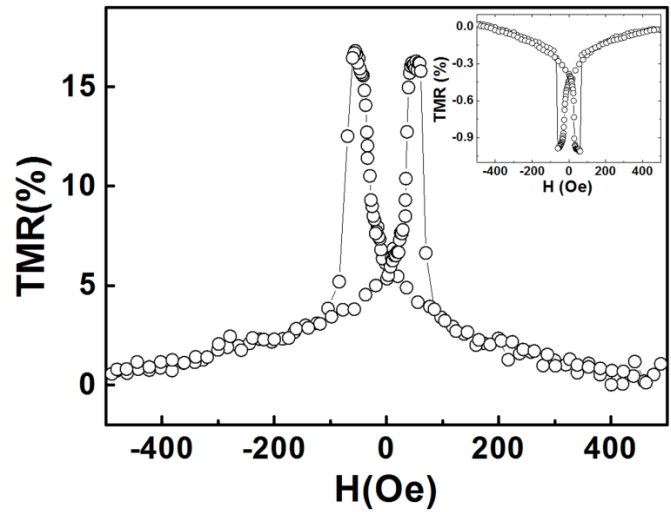

Fig.2: An example of the positive TMR at 4 K for the structure (in nm): 10 MgO / 4 $Co_{50}Fe_{50}$ / 1 LiF / 3 EuS / 10 Ti, the measurement voltage was 63 mV. The junction impedance is 120 MΩ (and 3.5 MΩ at room temperature). The inset shows an example of negative TMR of the same junction at 0.98 V.

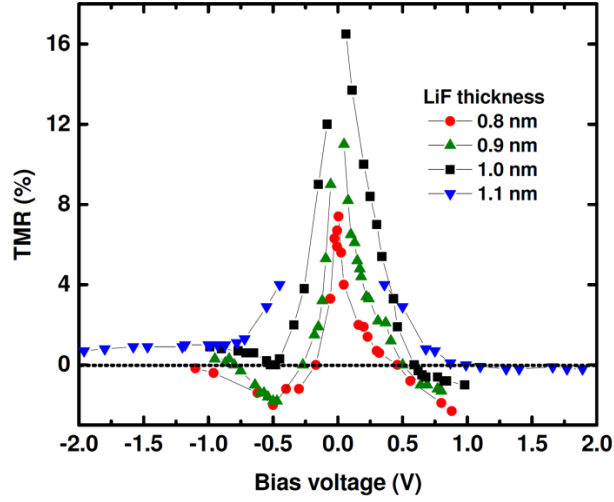

Fig.3: TMR bias voltage dependence for various thicknesses of LiF at 4 K. The EuS thickness was fixed at 3 nm. Some low voltage data were not collected due to the instruments' limitation for large impedance.

TMR inversion has been reported when the applied bias voltage gets higher than the barrier conduction bands [14,15]. However, to get above the conduction band of LiF (band gap 13.6 eV) is formidable, and to get above that of EuS requires > 1 V bias [11], thus in our system the TMR inversion cannot be attributed to quantum wells forming on top of the barrier conduction bands. Due to the very large junction impedance, single step tunneling is strongly suppressed and defect mediated hopping is promoted. This is also evidenced from the large temperature dependence of junction resistance. Thus the inverse TMR vs voltage is attributed to the energy and location of the localized defects states. Generally, for electrons matching certain energies, they could interact with certain defect levels inside the tunnel barrier, leading to inverse TMR from resonant tunneling [6]. As a result, the inversion only shows up when the applied bias voltage is large enough. In addition, the bias voltage also leads to a tilt of the barrier potential, therefore the exact location of the defect levels can also modify the onset voltage for the inversion. When the defect levels reach enough population, it is easier for the tunnel cur-rents to go through these levels rather than directly tunnel across the barrier. The spin current is no longer proportional to the DOS near the Fermi level of the electrodes. As we know, no TMR inversion has been observed in similar EuS junctions (coupled with $Al_2O_3$ or MgO), we can therefore safely conclude that our observed TMR inversion is originating from LiF. This is readily understandable because LiF is a compound not as stable as $Al_2O_3$ or MgO, and is prone to ion motion therefore defect formation in the growth process. On the other hand, as the LiF thickness increases, its band structure is more established and the impurity levels are more filled up or become deeply buried. As a consequence, the tendency of TMR inversion gradually disappears when LiF thickness increases and the bulk properties recover.

**Conclusion:**

In conclusion, we observed clear TMR inversion in LiF/EuS based quasi spin-filter tunnel junctions. The inversion is attributed to resonant tunneling through defect states of LiF, and the tendency of inversion quickly diminishes as the LiF layer becomes thicker and more perfected.

**Acknowledgement:**

This work was supported by the Natural Sciences and Engineering Research Council of Canada (NSERC) Discovery grant RGPIN 418415-2012, and the National Natural Science Foundation of China (NSFC) grant 11365009. The authors would also like to thank Industry Canada and Infinite Potential Inc. for providing the necessary infrastructures.

* Corresponding author: guo-xing.miao@uwaterloo.ca